\documentclass[prb,paper]{revtex4-2} 

\newcommand{\amcite}[1]{\textsuperscript{\cite{#1}}}

\usepackage{amsmath}  
\usepackage{amsfonts} 
\usepackage{graphicx} 
\usepackage{hyperref}

\newcommand{\beq}{\begin{equation}}
	\newcommand{\eeq}{\end{equation}}
\newcommand{\bqa}{\begin{eqnarray}}
	\newcommand{\eqa}{\end{eqnarray}}
\newcommand{\nn}{\nonumber}

\newcommand{\smallfrac}[2]{\mbox{$\frac{#1}{#2}$}}

\newcommand{\half}{\smallfrac{1}{2}}

\begin{document}

\title{Simple precession calculation for Mercury: a linearization approach}

\author{Michael J. W. Hall}
\affiliation{Theoretical Physics, Research School of Physics, Australian National University, Canberra ACT 0200, Australia}



\begin{abstract}
The additional precession of Mercury due to general relativity can be calculated by a method that is no more difficult than solving for the Newtonian orbit. The method relies on linearizing the relativistic orbit equation, is simpler than standard textbook methods, and is closely related to Newton's theorem on revolving orbits. The main result is accurate to all orders in $\tfrac{1}{c}$ for near-circular orbits.

\end{abstract}

\maketitle 

\section{Introduction} 

An important and early success of general relativity was to explain an anomalous precession of the perihelion of Mercury by $43^{\prime\prime}$ of arc per century,\amcite{history} making it highly desirable to derive this result in the classroom. Standard textook derivations typically either apply perturbation theory to the orbit equation\amcite{textbook,rindler,odonnell} or rely on an approximate factorisation of the relativistic energy equation\amcite{weinberg,hartle,hall} (formally related to Einstein's original calculation\amcite{history}). However, the first method requires finding a particular integral of an inhomogenous second-order differential equation (usually simply supplied to students) and extracting its nonperiodic component, while the second requires evaluating a nontrivial integral via nonobvious changes of variables (again usually supplied to students). 

Alternative derivations in the literature similarly tend to have challenging features for students in an introductory course. For example, a method based on the Runge-Lenz vector is very elegant but requires sophisticated mathematical machinery.\amcite{runge} 
Further, while approaches based on the small eccentricities of near-circular orbits are more elementary in character, they have required {\it ad hoc} assumptions about the typical radius\amcite{rowlands,nguyen}  and/or the angular momentum\amcite{mondragon} to obtain the correct result.

The purpose of this note is to point out a particularly simple method for calculating the precession in an introductory course, based on linearizing a quadratic term in the relativistic orbit equation. The resulting approximate orbit equation is as easy to solve as the Newtonian orbit equation, is able to estimate the precession to all orders in $\tfrac{1}{c}$ while avoiding {\it ad hoc} assumptions, and is related to Newton's method for estimating the precession of near-circular orbits.\amcite{newton,chandra} Necessary elements of Newtonian orbits are briefly reviewed in Sec.~\ref{sec:newt}; the linearization approach is presented in Sec.~\ref{sec:lin}; and underlying assumptions and comparisons with other methods are discussed in Sec.~\ref{sec:disc}.

\section{Newtonian orbits}
\label{sec:newt}

Newton's orbit equation for a particle moving in a given plane about a gravitational centre of mass $M$ is generally solved by writing it in the Binet form\amcite{textbook}
\beq \label{newton}
\frac{d^2u}{d\phi^2} + u = \frac{GM}{h^2},
\eeq
where $r=1/u$ is the orbital distance, $\phi$ is the orbital angle relative to the perihelion, $G$ is Newton's gravitational constant, and $h=r^2\tfrac{d\phi}{dt}$ is the conserved specific angular momentum. The general solution is found by adding the constant solution $u_0=\tfrac{GM}{h^2}$ to any solution of the homogenous equation $\tfrac{d^2u}{d\phi^2} + u =0$, to give 
\beq \label{newtsol}
\frac{1}{r} = u =  u_0  (1+e\cos \phi)
\eeq
for some constant $e$ (with $e\geq0$ to ensure perihelion at $\phi=0$). This represents a conic section, with $e<1$ for a bound orbit  (i.e., with $r<\infty$ at all times), corresponding to an ellipse with focus at the centre of mass and eccentricity $e$.

The parameters $u_0$ and $e$ in Eq.~(\ref{newtsol}) are fully determined by the perihelion distance $r_{\min}$ and aphelion distance $r_{\max}$ of the orbit, via 
$1/r_{\min}=u_0(1+e)$, $1/r_{\max}=u_0(1-e)$, yielding
\beq \label{u0e}
u_0 = \frac{r_{\max}+r_{\min}}{2r_{\max}r_{\min}},\qquad e = \frac{r_{\max}-r_{\min}}{r_{\max}+r_{\min}} .
\eeq
Thus observations of the perihelion and aphelion are sufficient to fix  the orbit.   One also has the useful relation
\beq \label{2a}
a:= \half(r_{\max} + r_{\min}) = \frac{1}{2u_0} \left(\frac{1}{1-e} + \frac{1}{1+e}\right) = \frac{1}{u_0(1-e^2)}.
 \eeq
Note that $a$ is the semi-major axis of an elliptical orbit.\amcite{latus}

\section{Precession of relativistic orbits}
\label{sec:lin} 

In general relativity, the motion of a body orbiting a much larger mass $M$ in a given plane, such as Mercury orbiting the Sun, is well-modeled by the equation for a geodesic orbit around a stationary black hole of mass $M$,\amcite{textbook,rindler,odonnell,weinberg,hartle,hall} 
\beq \label{einstein}
\frac{d^2u}{d\phi^2} + u = \frac{GM}{j^2} + \frac{3GM}{c^2} u^2,
\eeq
where $r=1/u$ and $\phi$ are the orbital distance and the angle relative to perihelion in Schwarzschild coordinates, $j=r^2\tfrac{d\phi}{d\tau}$ is the conserved specific angular momentum with respect to proper time, and $c$ is the speed of light in vacuum. The quadratic term on the right hand side has no analogue in the Newtonian orbit equation~(\ref{newton}) and is responsible for a non-Newtonian precession of bound orbits. The aim of this note is to give a simple method for estimating this precession.

The central idea is to linearize the quadratic term in Eq.~(\ref{einstein}) about the average value $\bar u$ of $u$, under the assumption that $(u-\bar u)^2\ll \bar u^2$ (equivalent to $|r-\bar u^{-1}|\ll r$). This is clearly a good approximation for near-circular orbits, and is closely related to Newton's theorem on revolving orbits, as discussed further below.  In particular, linearization gives
\begin{align} 
u^2 &= [\bar u + (u-\bar u)]^2 = \bar u^2 + 2\bar u (u-\bar u) + (u - \bar u)^2 \nn\\
&\approx \bar u^2 + 2\bar u (u-\bar u) \nn\\
&= 2\bar u u - \bar u^2,
\label{assump}
\end{align}
and substitution into Eq.~(\ref{einstein}) then yields the approximate orbit equation
\beq \label{approx}
\frac{d^2u}{d\phi^2} + K u = L
\eeq
for constants 
\beq \label{defkl}
K = 1-\frac{6GM\bar u}{c^2},\qquad L=\frac{GM}{j^2} - \frac{3GM}{c^2}\bar u^2 .
\eeq

Just as for the solution of the Newtonian orbit equation~(\ref{newton}), the general solution of Eq.~(\ref{approx}) is easily found by adding the constant solution, $u_0=L/K$, to any solution of the homogenous equation $\tfrac{d^2u}{d\phi^2} + Ku =0$, to give
\beq \label{relsol}
\frac{1}{r} = u = u_0(1 + e \cos \sqrt{K}\phi)
\eeq
for a suitable constant $e$, with $e\in[0,1)$ for bound orbits.  Noting that the cosine term varies between 1 and $-1$, it follows that the parameters $u_0$ and $e$ in Eq.~(\ref{relsol}) are related to the observed perihelion and aphelion distances of the orbit precisely as per Eqs.~(\ref{u0e}) and~(\ref{2a}) for Newtonian orbits.\amcite{latus}

However, whereas the Newtonian orbit in Eq.~(\ref{newtsol}) returns to perihelion at $\phi=2\pi$, the relativistic orbit in Eq.~(\ref{relsol}) returns to perihelion at $\sqrt{K}\phi = 2\pi$, i.e., at
\beq
\phi_p = \frac{2\pi}{\sqrt{K}} = \frac{2\pi}{\sqrt{1-\frac{6GM\bar u}{c^2}}} \approx 2\pi\left(1+\frac{3GM\bar u}{c^2}\right),
\eeq
to first order in $\tfrac{1}{c^2}$, using the approximation $(1-x)^{-1/2}\approx 1+\half x$.  Thus, in the relativistic case there is a precession of the perihelion per orbit by an amount
\beq \label{prec}
\Delta \phi = \phi_p-2\pi=2\pi\left(\frac{1}{\sqrt{K}}-1\right) \approx \frac{6\pi GM\bar u}{c^2} . 
\eeq

Finally, to determine the average value $\bar u$ in Eq.~(\ref{prec}), note that $\sqrt{K}\phi$ in Eq.~(\ref{relsol}) ranges over 0 to $2\pi$ from the initial perihelion to the next, so that the cosine term is zero on average, giving $\bar u=u_0$.  Hence, applying Eqs.~(\ref{u0e}) and~(\ref{2a}),
\beq \label{baru}
 \bar u = u_0 = \frac{r_{\max}+r_{\min}}{2r_{\max}r_{\min}} = \frac{1}{a(1-e^2)} ,
\eeq
and substitution into Eq.~(\ref{prec}) gives
\beq \label{precfin}
\Delta \phi \approx \frac{6\pi GM}{c^2}\frac{r_{\max}+r_{\min}}{2r_{\max}r_{\min}} = \frac{6\pi GM}{a(1-e^2)c^2}
\eeq
to first order in $\tfrac{1}{c^2}$. This is precisely the same expression for the relativistic precession calculated via standard methods in the literature,\amcite{textbook,rindler,odonnell,weinberg,hartle,hall} and for Mercury's orbit about the Sun evaluates to the well-known observed value of $43^{\prime\prime}$ of arc per century.

\section{Discussion}
\label{sec:disc}

The linearization method shows that the relativistic precession of Mercury's orbit may be obtained with no more mathematical sophistication than is required for solving the Newtonian orbit equation. In particular, no higher-order perturbative solutions\amcite{textbook,rindler,odonnell} or nontrivial integrals\amcite{weinberg,hartle,hall} need be evaluated. This makes the method very suitable for introductory courses in general relativity, in the form given above. For teachers and advanced students the additional points below may also be of interest.

First, it is worth noting that while Eq.~(\ref{precfin}) can be shown to be accurate  to first order in  $\tfrac{1}{c^2}$ and to all orders in $e^2$,\amcite{young} the general prediction of the  linearization method in Eq.~(\ref{prec}),
\beq \label{genprec}
\Delta \phi = 2\pi\left(\frac{1}{\sqrt{K}}-1\right) = 2\pi\left[\left(1-\frac{6GM}{a(1-e^2)c^2}\right)^{-1/2}-1\right] ,
\eeq
is, conversely, accurate to first order in $e^2$ and to all orders in $\tfrac{1}{c^2}$.  In particular, substituting  Eq.~(\ref{relsol}) into 
 the central assumption $(u-\bar u)^2\ll \bar u^2$ used in Eq.~(\ref{assump}), and recalling $u_0=\bar u$, gives the equivalent requirement
\beq \label{con}
e^2 \ll 1 
\eeq
for the validity of the linearization method, corresponding to near-circular orbits. For Mercury one has $e^2\approx 0.04$, so that the method accurately predicts the precession.

The simplicity and generality of Eq.~\ref{con} may be compared to the relatively complicated and restrictive condition required in a related approach for near-circular orbits by Lemmon and Mondragon,\amcite{mondragon,dragonfoot} which linearizes the orbit equation about $u_c:=j^2/(GM)$ and makes an {\it ad hoc} identification of the relativistic angular momentum $j$ with the angular momentum $h$ of a Newtonian orbit. This approach not only requires comparing relativistic and non-relativistic orbits having unequal perihelion distances but moreover, unlike Eq.~(14), is only accurate to first-order in $\tfrac{1}{c^2}$.

Second, there are  two weak constraints, $K>0$ and $L>0$, on the applicability of the linearization method to bound orbits,  implicit in the form of the general solution in Eq.~(\ref{relsol}). These constraints correspond to the existence of solutions with $0<r<\infty$, i.e., to physical bound orbits, and from Eqs.~\ref{defkl}) and~(\ref{baru}) can be written in the form
\beq
r_S:= \frac{2GM}{c^2} < \frac13 a(1-e^2),\qquad v_\parallel :=\frac{j}{a} < \frac{1-e^2}{\sqrt{3}} c,
\eeq
where $r_S$ is the Schwarzschild radius of the central mass $M$  and $v_\parallel$ is a measure of orbital speed. These constraints trivially hold for planetary orbits in our solar system (e.g., $r_S\approx 3$\,km for the Sun and $v_\parallel\approx 0.0002\,c$ for Mercury\amcite{angmom}).

Third, as mentioned in Sec.~\ref{sec:lin}, the linearization method is closely related to Newton's theorem on revolving orbits: if $r(\phi)$ describes an orbit corresponding to a central force per unit mass $f(r)$ and angular momentum $h$, then $r(k\phi)$ describes an orbit corresponding to a central force $f(r)+\tfrac{\alpha}{r^3}$ and angular momentum $H=h/k$, with $\alpha=(k^2-1)H^2$.\amcite{newton,chandra,deliseo} This theorem follows easily from the Newtonian orbit equation for the latter case,\amcite{textbook,chandra} 
\beq \label{U}
\frac{d^2u}{d\phi^2} + u = -\frac{1}{H^2u^2}[ f(u^{-1}) + \alpha u^3] = -\frac{1}{H^2u^2} f(u^{-1}) -\frac{\alpha}{H^2} u ,
\eeq
 which simplifies to
\beq
\frac{d^2u}{d(k\phi)^2} + u = -\frac{1}{h^2u^2} f(u^{-1}) ,
\eeq
with solution $u=1/r(k\phi)$ following from the definition of $r(\phi)$. The linear term on the righthand side of Eq.~(\ref{U}), corresponding to an inverse-cube force, plays the same formal role here as the linear term obtained via linearization in  Eq.~(\ref{assump}). Thus the simple connection between the forms of the elliptical and precessing orbits in Eqs.~(\ref{newtsol}) and~(\ref{relsol}) may be viewed as a special case of Newton's theorem on revolving orbits.

Further, Newton used his theorem to estimate the precession of orbits perturbed by an arbitrary additional force, by approximating the perturbing force by an inverse-cube force.\amcite{newton,chandra} However, while this is analogous to linearization of the quadratic term in Eq.~(\ref{einstein}), previous attempts to apply Newton's method to the relativistic precession of Mercury\amcite{rowlands,nguyen} have relied on an {\it ad hoc} assumption that is avoided by the linearization approach. In particular,  the relativistic orbit equation in Eq.~(\ref{einstein}) corresponds to a perturbing force $-\tfrac{3j^2GM}{c^2}\tfrac{1}{r^4}$ [{\it cf.} Eq.~(\ref{U})], which is approximated via the Taylor series expansion $\tfrac{1}{r}\approx\tfrac{1}{R} -\frac{r-R}{R^2}$ about some fixed value $r=R$ using $\tfrac{1}{r^4}=\tfrac{1}{r}\tfrac{1}{r^3}\approx \tfrac{2}{Rr^3}-\tfrac{1}{R^2r^2}$, i.e., by the sum of an inverse-cube term and an inverse-square term. The latter can be absorbed into the Newtonian gravitational force, and applying Newton's theorem for revolving orbits then leads to the corresponding precession\amcite{rowlands,nguyen}
\beq
\Delta \phi \approx \frac{6\pi GM}{Rc^2} 
\eeq
to first order in $\tfrac{1}{c^2}$. Comparing with Eq.~(\ref{precfin}), it follows that one should choose $R=a(1-e^2)$ to obtain the correct prediction. However, this choice is purely {\it ad hoc}. Rowlands simply makes this choice {\it ab initio}, without any physical motivation,\amcite{rowlands} while Nguyen instead chooses $R=a/(1-e^2)$, leading to a predicted precession of only $39^{\prime\prime}$ of arc per century for Mercury.\amcite{nguyen} An alternative choice, the average orbital distance
\beq
R=\bar r:= \frac{1}{2\pi}\int_0^{2\pi}  \frac{d\phi}{u_0(1+e\cos\phi)} = a\sqrt{1-e^2},
\eeq
again leads to an incorrect prediction.\amcite{timeav}  In contrast, the linearization method leads naturally and directly to the correct result in Eq.~(\ref{precfin}), and its generalization in Eq.~(\ref{genprec}).

It is concluded that the linearization method provides a simple and useful approach for calculating the general relativistic contribution to the precession of Mercury's orbit. The corresponding orbit equation~(\ref{approx}) is no more difficult to solve than the Newtonian orbit equation~(\ref{newton}), making the method particularly suitable for introductory courses; the precession is accurately estimated to all orders in $\tfrac{1}{c}$ for near-circular orbits (i.e., $e^2\ll1$); and the natural expansion about the average inverse radius $\bar u$ keeps the basic idea of Newton's method for calculating precession intact while removing the need for {\it ad hoc} assumptions.



\begin{thebibliography}{99}

\bibitem{history}  A. Einstein, ``Erkl\"arung der Perihelbewegung des Merkur aus der allgemeinen Relativit\"atstheorie'', \href{https://nms.kcl.ac.uk/gerard.watts/Einstein-1915.pdf}{Sitzungsber. K. Preuss. Akad. Wiss.} {\bf 47}, 831--839 (1915).
\bibitem{textbook} S. T. Thornton and J. B. Marion, {\it Classical Dynamics of Particles and Systems}, 5th edition (Brooks/Cole, Belmont, 2004), sections~8.4, 8.9, pp.~291--295, 312--316.

\bibitem{rindler} W. Rindler, {\it Relativity : Special, General, and Cosmological}, 2nd edition \href{https://www.oup.com.au/books/others/9780198567325-relativity}{(Oxford University Press, UK, 2006)}, section~11.9, pp.~241--245.

\bibitem{odonnell} P. J. O'Donnell, {\it Essential Dynamics and Relativity} \href{https://doi.org/10.1201/b17785}{(CRC Press, Boca Raton, USA, 2014)}, section~5.8, pp.~127--129.

\bibitem{weinberg} S. Weinberg, {\it Gravitational and Cosmology} (Wiley, New York, USA, 1972), section~8.6, pp.~194--201.

\bibitem{hartle} J. B. Hartle, {\it Gravity: an Introduction to Einstein's General Relativity} (Addison-Wesley, San Francisco, USA, 2003), pp.~216--217.

\bibitem{hall} M. J. W. Hall, {\it General Relativity: an Introduction to Black Holes, Gravitational Waves, and Cosmology}, \href{https://iopscience.iop.org/book/978-1-6817-4885-6}{(IOP Publishing, Bristol, UK, 2018)}, pp.~5-13--5-14.

\bibitem{runge} M. G. Stewart, ``Precession of the perihelion of Mercury’s orbit'', \href{https://doi.org/10.1119/1.1949625}{Am. J. Phys.} {\bf 73}, 730--734 (2005).

\bibitem{rowlands} P. Rowlands, ``A simple approach to the experimental consequences of
general relativity'', \href{https://doi.org/10.1088/0031-9120/32/1/020}{Phys. Educ.} {\bf 32}, 49--55 (1997). 

\bibitem{nguyen} H. D. Nguyen, ``Rearing its ugly head: the cosmological constant and
Newton’s greatest blunder'', \href{https://www.tandfonline.com/doi/abs/10.1080/00029890.2008.11920544}{Am. Math. Monthly} {\bf 115}, 415--430 (2008).

\bibitem{mondragon} T. J. Lemmon and A. R. Mondragon, ``Alternative derivation of the relativistic contribution to perihelic precession'', \href{https://aapt.scitation.org/doi/10.1119/1.3159611}{Am. J. Phys.} {\bf 77}, 890--893 (2009).

\bibitem{newton} I. Newton, {\it Mathematical Principles of Natural Philosophy}, trans. A Motte \& F. Cajori (University of California Press, Berkeley, USA, 1962), Book I section IX, pp. 135--147.

\bibitem{chandra} S. Chandrasekhar, {\it Newton's Principia for the Common Reader} (Clarendon, Oxford, UK, 1995), pp. 184--199.



\bibitem{latus} Note also in both Eqs.~(\ref{newtsol}) and~(\ref{relsol}) that $1/u_0$ is equal to the orbital distance at the angle halfway from perihelion to aphelion (corresponding to the semilatus rectum for elliptical orbits). 


\bibitem{young} K.-H. Lo, K. Young, and B. Y. P. Lee, ``Advance of perihelion'', \href{https://doi.org/10.1119/1.4813067}{Am J. Phys.} {\bf 81}, 695--702 (2013).

\bibitem{dragonfoot} The self-consistency condition in Eq.~(31) of Ref.~\cite{mondragon} is
$e+3(\tfrac{GM}{jc})^2(e+2)\ll 1$.

\bibitem{angmom} Planetary data is available at, e.g., the NASA website \url{<https://history.nasa.gov/SP-345/p16.htm>} (Accessed on May 2022).


\bibitem{deliseo} M. M. D'Eliseo, ``The first-order orbital equation'', \href{https://doi.org/10.1119/1.2432126}{Am. J. Phys.} {\bf 75}, 352--355 (2007).


\bibitem{timeav} Similarly, choosing $R$ to be the time average of $r$ between successive perihelia $P$ and $P'$ and using $r^2\tfrac{d\phi}{d\tau}=j$ gives $R=\int_P^{P'} r\, d\tau/\int_P^{P'} d\tau=\int_0^{2\pi} \tfrac{d\phi}{[u_0(1+e\cos\phi)]^3}/\int_0^{2\pi}\tfrac{d\phi}{[u_0(1+e\cos\phi)]^2}=a(1+\half e^2)$, thus also leading to an incorrect prediction.






%
%
%
%
%
%
%
%
%
%
%
%
%
%
%
%

\end{thebibliography}
\end{document}